\documentclass[aps,pra,floatfix,superscriptaddress,twocolumn,nofootinbib,showkeys]{revtex4}
\usepackage{graphicx}
\usepackage{subfig}
\usepackage{amssymb,amsmath,amsfonts,mathrsfs,eufrak,fancyhdr}
\usepackage{hyperref}
\usepackage{color}
\usepackage{makeidx,endnotes}
\usepackage{showidx}
\usepackage{theorem,eepic}
\usepackage{epsf,epsfig}
\usepackage{braket}
\usepackage{enumitem}
\usepackage{float}

\usepackage[utf8]{inputenc}

\def\<{\langle}
\def\>{\rangle}

\def\sH{\mathscr{H}}

\def\Tr{\operatorname{Tr}}

\begin{document}
\title{Quantum walk transport on carbon nanotube structures}
\author{J. Mare\v s, J. Novotn\'y, I. Jex}
\affiliation{Department of Physics, Faculty of Nuclear Sciences and Physical Engineering, Czech Technical University in Prague, B\v rehov\'a 7, 115 19 Praha 1 - Star\'e M\v esto, Czech Republic}
\date{\today}
\begin{abstract}
We study source-to-sink excitation transport on carbon nanotubes using the concept of quantum walks. In particular, we focus on transport properties of Grover coined quantum walks on ideal and percolation perturbed nanotubes with zig-zag and armchair chiralities. Using analytic and numerical methods we identify how geometric properties of nanotubes and different types of a sink altogether control the structure of trapped states and, as a result, the overall source-to-sink transport efficiency. It is shown that chirality of nanotubes splits behavior of the transport efficiency into a few typically well separated quantitative branches. Based on that we uncover interesting quantum transport phenomena, e.g. increasing the length of the tube can enhance the transport and the highest transport efficiency is achieved for the thinnest tube. We also demonstrate, that the transport efficiency of the quantum walk on ideal nanotubes may exhibit even oscillatory behavior dependent on length and chirality.
\end{abstract} 


\maketitle

%
%
%
Discrete coined quantum walks (CQWs) have become a standard tool in studying transport phenomena in the quantum domain \cite{review}. Over the last two decades, quantum walker's behavior has been analyzed in the context of recurrence phenomena \cite{Stefanak2008,Werner2013}, state transfer \cite{Tanner2009,Skoupy2016}, speed of wave packet propagation \cite{Aharonov2001}, hitting times \cite{Magniez}, or e.g. topological phenomena \cite{Asboth2012}. Investigations, aiming first at the quantum walker on the line, have gradually broadened the scope of their interest to different graph geometries like e.g. cycles \cite{Aharonov2001}, hypercubes \cite{Moore2002,Portugal2008}, trees \cite{Segawa2009}, honeycombs \cite{Lyu2015,Kendon2016}, spidernets \cite{Segava2013} or fractal structures \cite{Lara2013} (for more see review \cite{review}). 

Analysis of complex graph structures has revealed that if vertices with degree at least three are present, the walker's dynamics may exhibit localisation originating from the presence of so-called trapped states \cite{Inui2004,inui:psa,inui:grover1,miyazaki,watabe,falkner,machida}. They appear in various quantum systems and are known, in a different context, also as localized invariant or dark states \cite{Fleischhauer2000,Poltl2009,Creatore2013,Mendoza2013}. These are eigenstates of the given dynamics, whose support does not spread over the whole position space. Due to that, the initial states having overlap with the trapped states can not fully propagate through the medium and the efficiency of quantum transport may be significantly reduced \cite{Rebentrost2009,Chin2010}. 

On the other hand, trapped states were found to be fragile with respect to certain decoherence mechanisms arising in the presence of random external perturbations \cite{Caruso2009}. When the quantum walker moves on a changing graph whose edges are randomly and repeatedly closed and open again, we arrive at so-called dynamically percolated coined quantum walks (PCQWs) \cite{asymptotic1,asymptotic2} capable to destroy some trapped states of the original non-percolated CQW \cite{assisted_transport}. Recently, it was shown how the underlying geometry of Grover PCQW controls the structure of walker's trapped states \citep{theory}. The detailed analytical recipe for constructing the basis has essentially contributed to identify interesting counterintuitive quantum transport phenomena \citep{transport_effects}. In particular, studying Grover PCQWs on the ladder graph and Cayley trees demonstrated that by increasing the transport distance or adding redundant blind branches one can eventually enhance excitation source-to-sink quantum transport. Since all trapped states of PCQWs are trapped states of their non-percolated versions, these counterintuitive transport phenomena apply to CQWs as well.
\begin{figure}[th]
	\centering
	\includegraphics[width=160 pt]{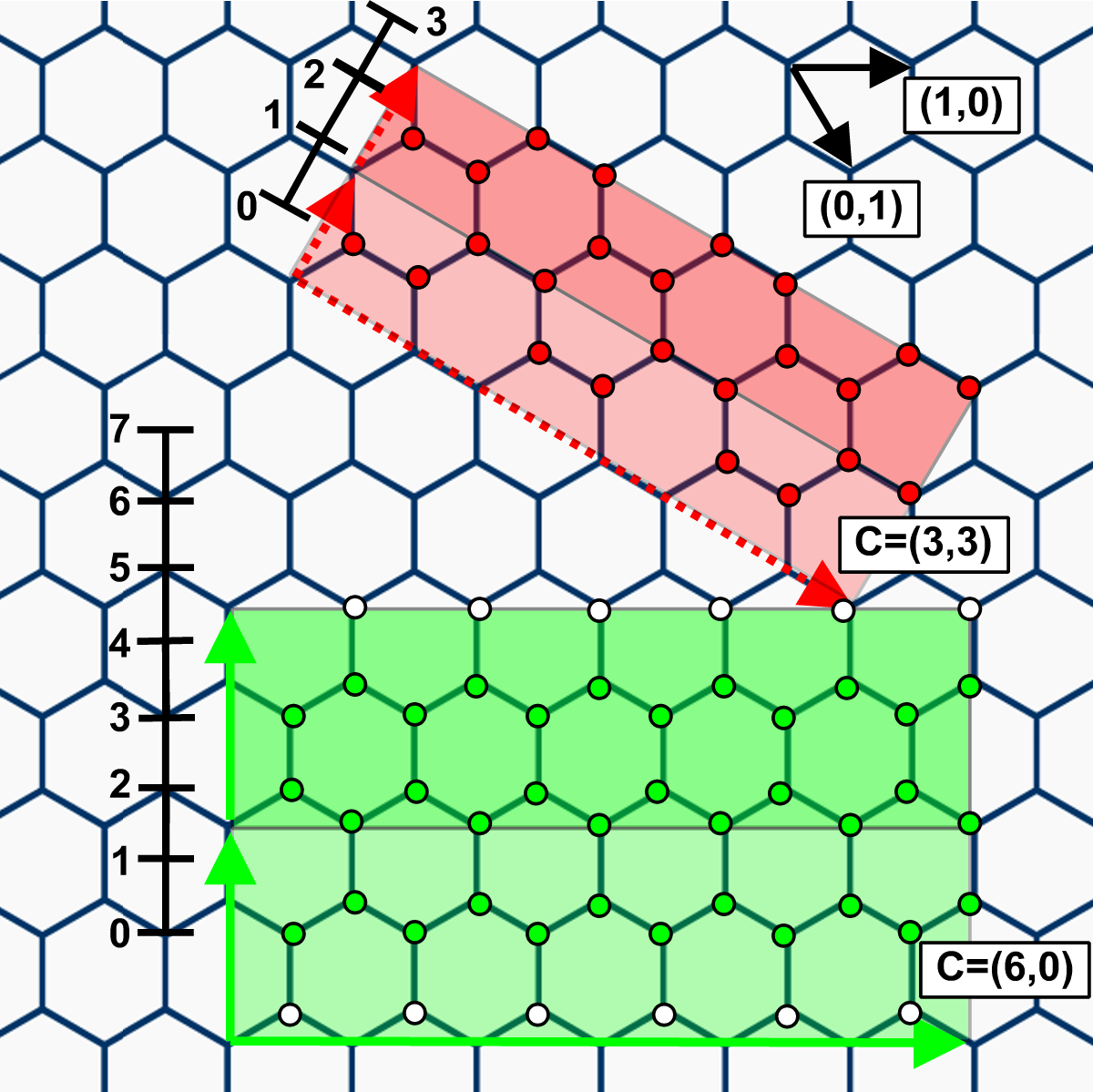}
	\caption{(color online) Schematic explanation of the chirality of nanotubes: the basis vectors $(1,0)$ and $(0,1)$, chirality vectors of two example tubes: $(6,0)$ chirality vector as a solid green arrow and $(3,3)$ vector as a dotted red arrow with the corresponding orthogonal vectors determining two basal length segments of tubes. Vertices contained in tubes are shown (identified by circles). Note that in the case of $(n,0)$ tubes we subsequently cut out the vertices of degree 1 (white) before using the structure for a quantum walk. Tube length scale is shown using the distance of neighboring nodes as the unit of length.}
	\label{fig:chirality}
\end{figure}

In this paper we explore transport properties of quantum walks on realistic carbon nanotube structures. Related to the above, we focus primarily on excitation source-to-sink quantum transport for Grover PCQWs and the analytically obtained findings are subsequently numerically checked for Grover CQWs. Recently, the speed of excitation transport for the Grover CQW on carbon nanotubes was numerically studied for one particular initial state, which is the only one that exhibits full source-to-sink transport \cite{Kendon2016}. Unfortunately, numerical simulations do not constitute an efficient tool for unravelling a delicate skein of different regimes of walker's behavior, when no analytical insight is available. Here we show that carbon nanotubes share a similar structure of trapped states with the ladder graph. Consequently, these nanotubes, despite their significantly higher complexity compared to the ladder, also exhibit the enhancement of the source-to-sink transport by extending the transport distance. Moreover, the shape of important trapped states attached to open nanotube ends is controlled by a characteristic structure parameter of carbon nanotubes - the chirality. This geometrical aspect of carbon nanotubes splits transport efficiency into different quantitative branches whose mutual gaps can be estimated even from these non-orthogonal trapped states. In PCQWs, the diameter of the tube further generates a subtle quantitative separation within these branches following the rule that thinner tubes exhibit a better transport. In CQWs this behavior is more complex but also here the thinnest nanotube shows the best transport. Moreover, due to a richer family of trapped states we can even identify oscillatory behavior of the overall transport efficiency for a certain type of the sink.

Let us introduce a general model of walker's dynamics refined for the case of carbon nanotubes. We adopt the definition of a coined quantum walk presented in \citep{theory} as it is able to capture simultaneously the complexity of carbon nanotube structures and random disturbances of the underlying medium (percolation). 

A quantum walk is defined on a pair of finite graphs. The first undirected structure graph $G(V,E)$ describes the geometry of the underlying walker's medium. The set of discrete vertices (nodes) $V$ represents all possible walker's positions on the nanotube. The walker can travel among vertices in both directions along undirected edges from $E$. In our case the structure graph is a coiled strip of the honeycomb lattice. The way how we cut out the strip of the honeycomb lattice imprints basic shape to the carbon nanotube  and it is given by the chirality vector $(m,n)$, where $m$ and $n$ are positive integers. The numbers $m$ and $n$ are coefficients in a non-orthogonal basis defined on a honeycomb lattice as shown in figure \ref{fig:chirality}. The circumference of the tube is given by associating the beginning of the chirality vector with its end-point. The orthogonal vector represents the axial direction of the tube. For any chirality there is a basic length segment, which is repeated in the axial direction.

\begin{figure}
	\centering
	\includegraphics[width=180 pt]{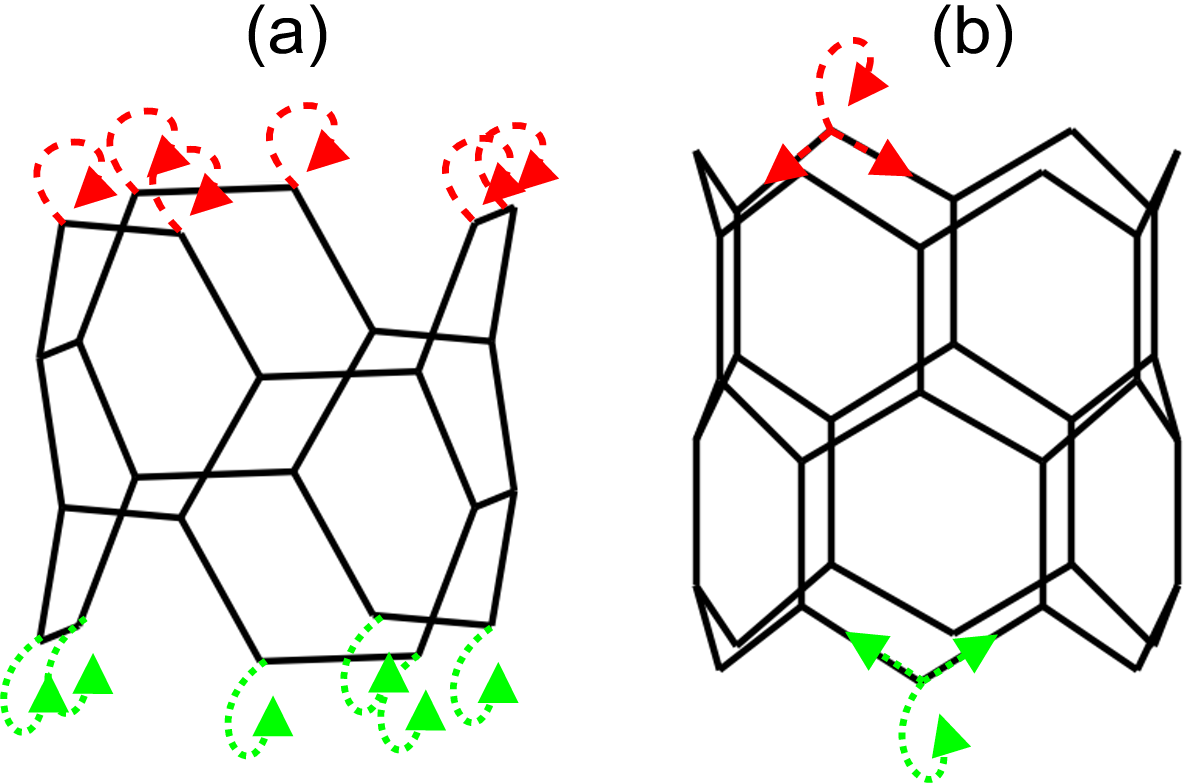}
	\caption{Examples of the initial subspace (green dotted arrows) and sink subspace (red dashed arrows) variants: loops type shown on a $(3,3)$ tube (a) and one-vertex type shown on a $(6,0)$ tube (b).}
	\label{fig:tubes_sink_init}
\end{figure}

In this work we study transport properties for two distinct classes of nanotubes, namely $(n,n)$ "armchair" tubes (Fig. \ref{fig:tubes_sink_init} (a)) and $(n,0)$ "zig-zag" tubes (Fig. \ref{fig:tubes_sink_init} (b)).  Nevertheless, the general results are mostly valid for tubes with arbitrary chirality. The structures are generated using TubeGen 3.4 \citep{tubegen}.

\begin{figure}
	\centering
	\includegraphics[width=140 pt]{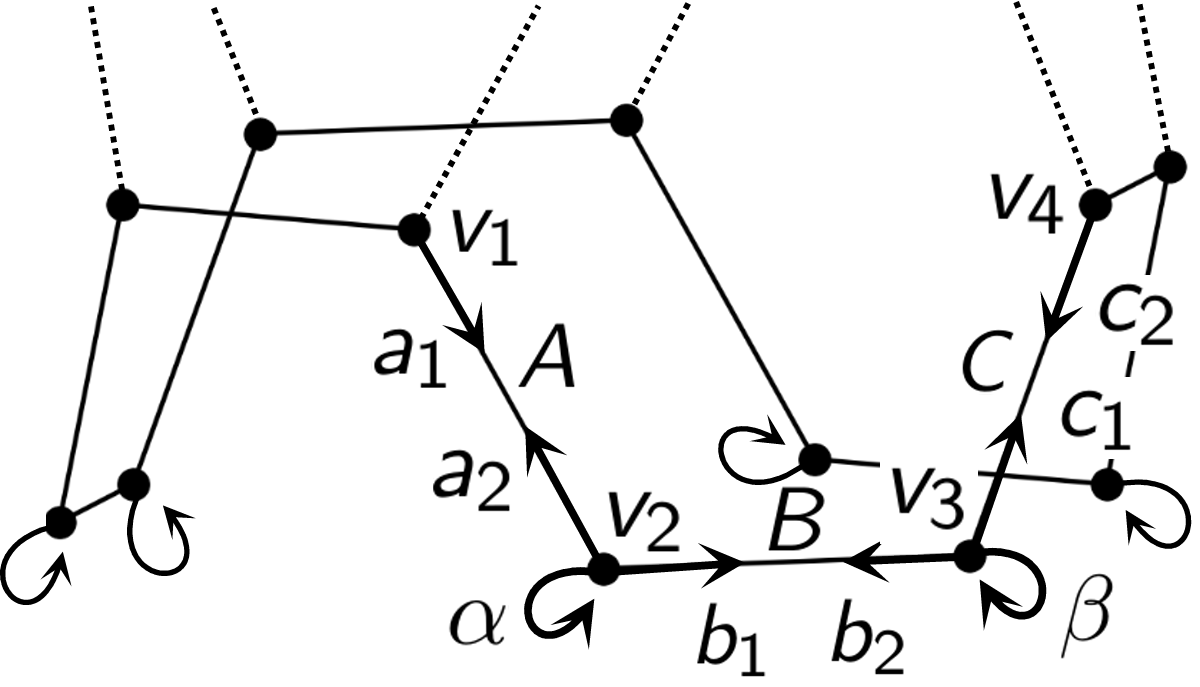}
	\caption{Example of both, structure and state, graphs demonstrated on the end of the $(3,3)$ armchair nanotube. Four vertices $v_1$ to $v_4$ are connected by three undirected edges $A$, $B$ and $C$ from the structure graph. For all of them we have three pairs of directed arcs $a_1$, $a_2$, $b_1$, $b_2$, $c_1$ and $c_2$ and additionally two self-loops $\alpha$ and $\beta$ from the state graph. The Hilbert space of the corresponding walk is $\sH=\mathrm{span}(\ket{a_1},\ket{a_2},\ket{b_1},\ket{b_2},\ket{c_1},\ket{c_2},\ket{\alpha},\ket{\beta},\ldots)$ and e.g. the vertex subspace $\sH_{v_2}=\mathrm{span}(\ket{a_2},\ket{b_1},\ket{\alpha})$.}
	\label{fig:structure_and_state}
\end{figure}

For the description of walker's state space and evolution we also introduce a directed state graph $G^{(d)}(V,E^{(d)})$. The set of vertices is the same as in the structure graph and we assign to each undirected edge $e = \{v_1,v_2\}\in E$ in the structure graph $G$ two directed arcs $e_1=(v_1,v_2), e_2=(v_2,v_1) \in E_p^{(d)}$ connecting the vertices in both directions. Therefore, each inner vertex of the state graph associated with the nanotube has three outgoing and three incoming directed edges. In order to make the whole state graph 3-regular, we add self-loops $e_l \in E_l^{(d)}$ beginning and ending in the same vertex which belongs to one of the open ends of the tube. An example of both graphs is depicted in Fig. \ref{fig:structure_and_state}. Each directed edge (including self-loops) $(v_1, v_2) \in E_l^{(d)}$ is then associated with a base state of an orthonormal basis and represents the walker standing in the vertex $v_1$ and facing towards $v_2$. Further, for every vertex $v\in V$ we denote the vertex-subspace of states corresponding to edges beginning in $v$ as $\sH_v$. Again, see the example of the notation introduced above in Fig. \ref{fig:structure_and_state}. The whole walker's Hilbert space $\sH$ can be written as
\begin{align}
\sH = \mathrm{span}\left(\ket{e^{(d)}} | e^{(d)} \in E^{(d)} \cup  E_l^{(d)} \right) = \bigoplus_{v \in V} \sH_v.
\end{align}

Each step of discrete CQW evolution is realized by a subsequent application of three operators. First, the reflecting shift operator $R$ (also called flip-flop) displaces the walker by swapping pairs of amplitudes corresponding to paired arcs $R\ket{(v_1,v_2)}= \ket{(v_2,v_1)}$ and it leaves the amplitudes corresponding to self-loops unchanged. Then the Grover coin operator $C$ acts locally in each vertex subspace as the Grover matrix
\begin{align*}
G_3 &= 
\frac{1}{3}\left[
\begin{array}{rrr}
 -1 & 2 & 2 \\
 2 & -1 & 2 \\
 2 & 2 & -1 \\
\end{array}
\right].
\end{align*} 
The simultaneous choice of the Grover coin and the reflecting shift operator results in a rich family of trapped states of CQWs and PCQWs. Their general structure as well as the effect of the order of $C$ and $R$ is discussed in \citep{theory}. Finally, at the end of each step the part of the walker's wave function which reached the sink is taken away. Our aim is to study excitation transport from an initial region to a target region represented by the sink. For a given walk the sink is defined as a set of directed edges in the state graph and the span of the associated states constitutes the sink subspace  $\sH_s$. Note that this definition includes the setting with no edge in the sink set, which is referred as a quantum walk without sink. We introduce the projector onto the sink subspace $\Pi_{\sH_s}$ and its complement $\Pi = I-\Pi_{\sH_s}$. Thus, each step is finished by a projective measurement of the walker's wave function on the complement of the sink subspace. Overall, one step of the CQW maps the walker from a state $\rho(t)$ to the state
\begin{equation}
\label{step_percolated_sinked}
\rho(t+1)=\Pi C R \rho(t) R C \Pi.
\end{equation}

The PCQW incorporates into its dynamics the so called dynamical percolation \citep{asymptotic1}. In each its step a subset of open edges $K \subset E$ is chosen randomly. Only those can be traversed by the walker. The remaining edges are closed in both directions. This results in a different reflecting shift operator $R_K$ for every configuration of open edges $K$. The arcs on closed edges are treated as loops - their amplitudes are not swapped. As the actual configuration of open edges is not under control, the resulting step of the PCQW evolution takes into account all of them and reads
\begin{equation}
\label{step_percolated}
\rho(t+1)=\sum_{K\subset E}\pi_K \Pi C R_K \rho(t) R_K C \Pi,
\end{equation}
where $\pi_K$ denotes the probability distribution of these configurations.

The evolution of CQWs and PCQWs is not, in general, trace preserving and $p(t) = \Tr\left( \rho(t) \right)$ expresses the probability of the walker being still away from the sink.  In contrast to the classical random walk the quantum walker is able to avoid the sink indefinitely and we can quantify the overall ability of the quantum system to transport the excitation initiated in the state $\rho(0)$ into the sink by asymptotic transport probability (ATP)
\begin{equation}
\label{def_efficiency}
q(\rho(0)) = 1-\Tr\left(\lim_{t \rightarrow +\infty} \rho(t)  \right).
\end{equation}
The complement $p=1-q$ is the trapping probability. It is due to the presence of trapped states whose overlap with the walker's initial state determines the ATP \citep{theory,assisted_transport}. The structure of trapped states is determined not only by chosen nanotube parameters (its length and chirality) but also by the choice of the sink. Indeed, trapped states of CQWs or PCQWs with a sink are all trapped states of the corresponding quantum walk without sink, which additionally have zero overlap with the sink subspace. They are called sr-trapped states ("sink-resistant") to distinguish them from trapped states of the corresponding quantum walk without sink. A different choice of the sink results in a different selection of sr-trapped states from the set of trapped states.

Our intention is to study ATP for different choices of sink and initial state $\rho(0)$. Similarly to the sink subspace we choose a subset of directed edges to define the initial subspace $\sH_i$. Once the initial subspace $\sH_i$ is chosen, each walker's initial state has non-zero amplitudes only in the initial subspace. We primarily study ATP averaged over all possible initial states from a given initial subspace, which due to linearity can be expressed as $\overline{q}= q(\overline{\rho})$ with $\overline{\rho}$ being the maximally mixed state on the initial subspace. For convenience of the argument we place the tube vertically with the walker initiated at the bottom and the sink at the top. In order to obtain a representative picture of quantum transport on nanotubes we use two variants of the sink and initial subspaces. In the "one-vertex"~variant (Fig. \ref{fig:tubes_sink_init} (b)) the subspace coincides with the vertex subspace of one chosen end vertex, e.g. $\sH_i=\sH_{v_2}=\mathrm{span}(\ket{a_2},\ket{b_1},\ket{\alpha})$ in Fig. \ref{fig:structure_and_state}. The second is "loops"~variant (Fig. \ref{fig:tubes_sink_init} (a)), where the subspace is formed by all loop states at one end of the tube, e.g. $\sH_i=\mathrm{span}(\ket{\alpha},\ket{\beta},\ldots)$ in Fig. \ref{fig:structure_and_state}. Thus we consider that the walker can enter (leave) the tube either at one chosen bottom (top) vertex or via bottom (top) loops. These two variants define four different regimes of transport, which we refer to as vertex-to-vertex, vertex-to-loops, loops-to-vertex, and loops-to-loops transport.

\begin{figure}
	\centering
	\includegraphics[width=240 pt]{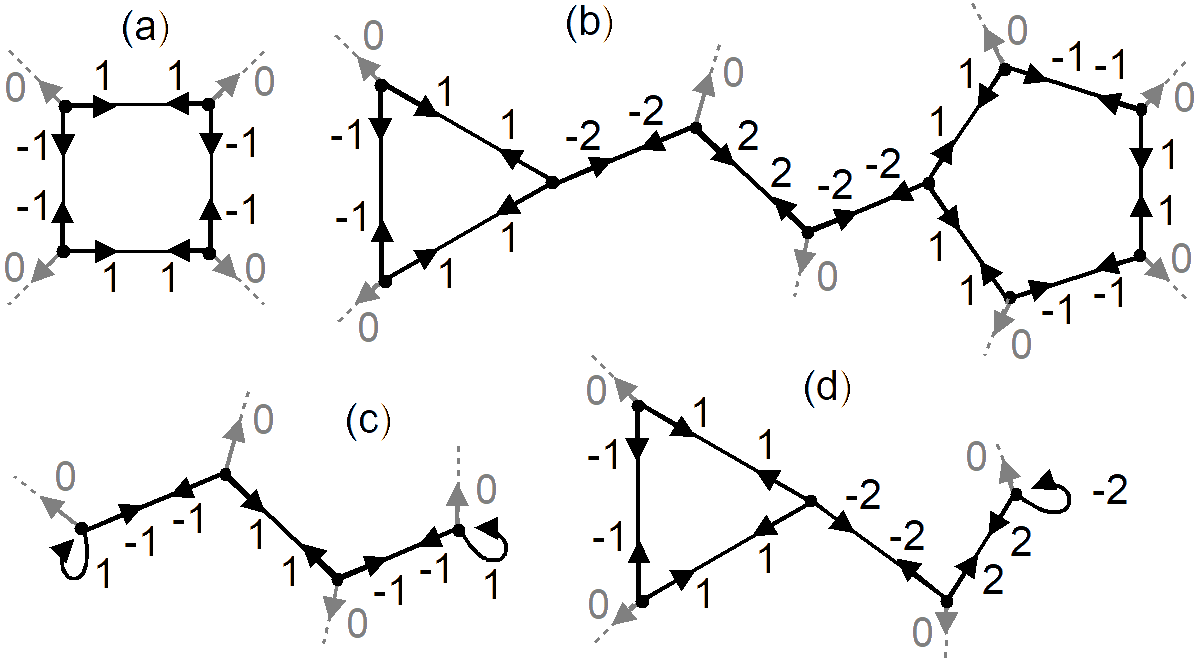}
	\caption{The four types of trapped eigenstates: (a) A-type state on an even-edged face, (b) B-type state connecting two odd-edged faces by a connecting path, (c) C-type state connecting two loops by a path and (d) D-type state connecting an odd-edged face and a loop by a path. Dashed lines represent continuation of the graph where all vector elements of the given trapped state are zeros.}
	\label{fig:trapped_construction}
\end{figure}

Let us analyze how the geometry parameters of nanotubes and different transport regimes of affect ATP behavior. This requires to construct basis of sr-trapped states for each investigated setting.
To proceed, we exploit a general recipe given in \citep{theory} allowing to construct a basis of trapped states for the reflecting Grover percolated coined quantum walk on an arbitrary simple planar 3-regular state graph. 
All trapped states correspond to eigenvalue $-1$ and they form a subspace of the dimension $N = 2\#V-\#E$. A basis of this subspace can be constructed from four types of trapped states shown in Fig. \ref{fig:trapped_construction}. For detailed description of these states see \citep{theory}. 

The construction of the basis relies on faces, which are defined in planar graphs. A planar embedding of our tube can be obtained by first transforming it from a cylinder to a truncated cone by sufficiently extending the top end and then projecting it in the axial direction into the plane. The former top end now forms the so called outer face - the rest of the plane outside of the graph. Further, there are all the hexagonal faces and the last face originating from the bottom end of the tube. The "bottom"~and "top"~faces have $4n$ and $2n$ edges in $(n,n)$ and $(n,0)$ tubes respectively and the number is even also for all other chiralities \footnote{ The base vectors shown in Fig. \ref{fig:chirality} always connect vertices separated by two edges. Since the lattice is hexagonal, also any other path connecting the origin and the site given by the chirality vector consists of even number of edges. (Always, $k$ edges can be replaced by $6-k$ others in the path.) As the bottom and top faces originate from such paths, they have even number of faces regardless of the particular chirality vector.}. We find all faces of the planar embedding are even-edged. Therefore, to construct basis of trapped states we need only A-type and C-type states here. 

\begin{figure}
	\centering
	\includegraphics[width=190 pt]{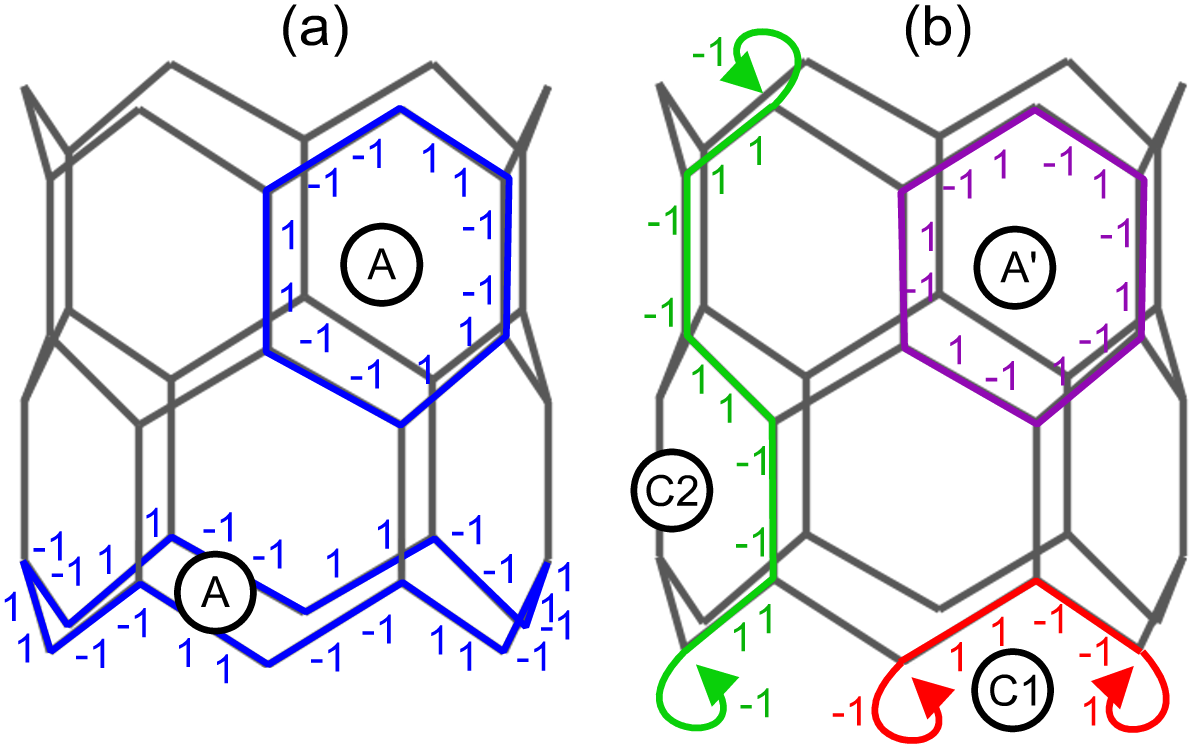}
	\caption{The three types of trapped states needed for PCQWs on nanotube structures: (a) A-type states on even-edged faces (blue), (b) C1-type "short path"~state (red) and C2-type "connecting path"~state (green), and an additional A'-type eigenstate for non-percolated CQWs (purple) also in (b) all shown on a $(6,0)$ nanotube structure with two length segments.}
	\label{fig:trapped_states_both}
\end{figure}

According to the recipe we should first include into the basis one A-type state for every face of our graph (except the outer face) and add C-type states connecting one chosen and fixed loop to all the other loops by arbitrary connecting paths. Nevertheless, we can create a more convenient basis using different linear combinations of C-type states in our particular case. We use C1-type states as shown in Fig. \ref{fig:trapped_states_both} (b) connecting all pairs of closest loops on ends of the tube except one on every end. Finally we add one C2-type "connecting state". One can see that for a tube with $n$ loops on every end we just replace $2n-1$ trapped states with the same number ($(n-1)+(n-1)+1$) of others, which are linear combinations of the original ones. Note, the two left out C1-states and the loops in the connecting state can be chosen arbitrarily. When we choose trapped states we try to avoid their overlap with the sink as much as possible (one vertex sink). This simplifies the subsequent selection of sr-trapped states and their basis, which have zero overlap with the sink subspace as detailed in \citep{theory}. We can just remove the trapped states overlapping with the sink subspace.
The obtained basis is typically non-orthogonal and has to be orthonormalized (possibly numerically). 

\begin{figure}
	\centering
	\includegraphics[width=240 pt]{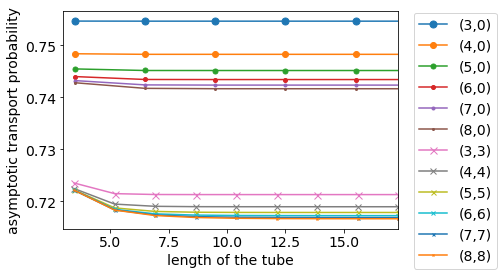}
	\caption{The average ATP for the PCQW in the vertex-to-loops transport regime for different chiralities and lengths of the tube.}
\label{fig:perc_avg_sloops_ione}
\end{figure}

\begin{figure}
	\centering
	\includegraphics[width=240 pt]{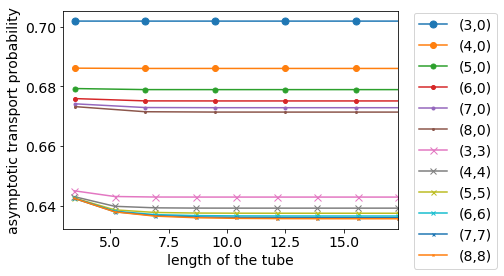}
	\caption{The average ATP for the PCQW in the loops-to-loops transport regime for different chiralities and lengths of the tube.}
\label{fig:transport_decrease}
\end{figure}

The trapping effect of A-type states is particularly well illustrated in vertex-to-loops and loops-to-loops transport regimes, in which the C2-type state is not sr-trapped. Necessary orthonormalization of the A-type sr-trapped states can make all of them overlapping with the initial state. Thus, if we extend the tube the ATP is pushed down slightly by new A-type trapped states. Nevertheless, the magnitude of this effect drops exponentially and ATP becomes constant as seen in Fig. \ref{fig:perc_avg_sloops_ione} and Fig. \ref{fig:transport_decrease}.

\begin{figure}
	\centering
	\includegraphics[width=240 pt]{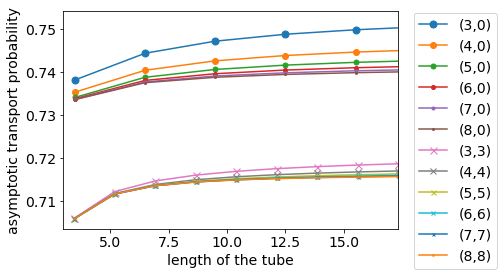}
	\caption{The average ATP for PCQW in the vertex-to-vertex transport regime for different chiralities and lengths of the tube.}
	\label{fig:perc_avg_sone_ione}
\end{figure}

\begin{figure}
	\centering
	\includegraphics[width=240 pt]{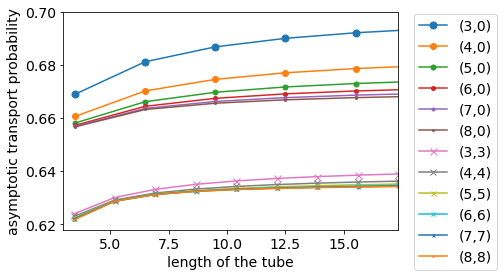}
	\caption{The average ATP for PCQW in the loops-to-vertex transport regime for different chiralities and lengths of the tube.}
	\label{fig:transport_increase}
\end{figure}

\begin{figure}
	\centering
	\includegraphics[width=240 pt]{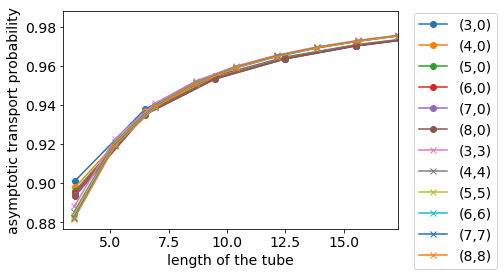}
	\caption{The ATP for PCQW in the loops-to-vertex transport regimes numerically maximized for each setting of length and chirality.}
	\label{fig:strong_trapping}
\end{figure}

On the other hand, in vertex-to-vertex and loops-to-vertex transport regimes the C2-type state is sr-trapped and we may observe that the ATP actually increases with extension of the tube, see Fig. \ref{fig:perc_avg_sone_ione} and Fig. \ref{fig:transport_increase}. Surprisingly, the walker is more likely to traverse a longer tube than a shorter one. The explanation is surprisingly simple too. An extension of the tube stretches the C2-type state, increases its number of vector elements and due to normalization its overlap with the initial subspace decreases. The effect was already reported in CQWs and PCQWs on much simpler geometry of the ladder graph \cite{transport_effects}. Here we show that the effect can be observed for the physically relevant, however more complex, structures of carbon nanotubes. Please note that in Fig. \ref{fig:perc_avg_sone_ione} and Fig. \ref{fig:transport_increase} we depict the averaged ATP. In Fig. \ref{fig:strong_trapping} we present the same effect, here with a significantly higher increase of ATP, for numerically obtained initial states maximizing ATP for each depicted setting. Moreover, Fig. \ref{fig:strong_trapping} demonstrates also another interesting effect similar to the so called strong trapping effect \cite{Kollar2015}. Indeed, the presence of the C2-type trapped state in the loops-to-vertex transport regime excludes existence of an initial state exhibiting the complete transport. Yet, as the C2-type trapped state recedes, the ATP approaches one for the maximal-transport states.

In the following we focus on geometric parameters of tubes, the chirality and the diameter, whose impact on transport properties of tubes has not been discussed yet. From all figures \ref{fig:perc_avg_sloops_ione}, \ref{fig:transport_decrease}, \ref{fig:perc_avg_sone_ione} and \ref{fig:transport_increase} one can see a clear quantitative splitting of the average ATP for tubes with chiralities $(n,0)$ and $(n,n)$. 
Its explanation arises from the manner in which the chirality entirely controls the structure of C1-type trapped states and weakly also the structure of A-type trapped states. In particular, all the C1-type states in $(n,0)$ tubes have six alternating elements 1,-1 (see Fig. \ref{fig:trapped_states_both}) and therefore are normalized by $\frac{1}{\sqrt{6}}$ whereas half of C1-type states in $(n,n)$ tubes have four and half of them has eight alternating elements 1,-1 properly normalized. Based on that we can roughly estimate the difference between averaged ATP of both chiralities. As an example we provide a reliable estimate for the loops-to-loops transport regime, for which the loops initial subspace has no overlap with A-type trapped states. Indeed, while for chirality $(2n,0)$ the averaged ATP can be estimated as $q_1 \approx 1-2n \left((1/(2n) (1/6 + 1/6)\right) = 2/3$, for chirality $(n,n)$ we obtain estimation $q_2\approx 1-2n \left((1/(2n) (1/4 + 1/8)\right) = 5/8$. In words, all $2n$ loops contribute with the same overlap between the corresponding component of the maximally mixed state and two normalized C1-type trapped states attached to one loop.

From figures \ref{fig:perc_avg_sloops_ione}, \ref{fig:transport_decrease}, \ref{fig:perc_avg_sone_ione},  and \ref{fig:transport_increase}, it is also apparent that the average ATP slightly differs for different diameters of the tube and better transport is surprisingly achieved for thinner tubes. The effect is present in all studied transport regimes. It is due to the dominant feature that more trapped states are added with increasing diameter, which results in a higher trapping probability. However, this effect diminishes for large diameters, as these new trapped states have gradually very low overlaps with loops and vertex initial subspaces.

\begin{figure}
	\centering
	\includegraphics[width=190 pt]{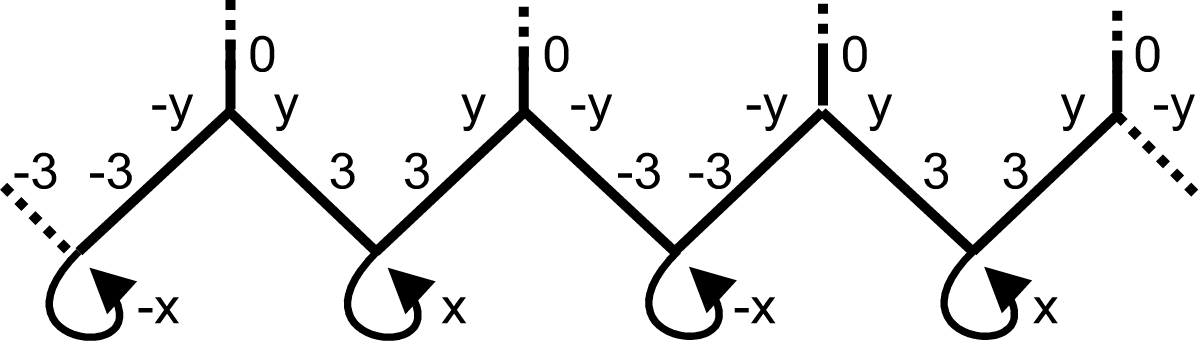}
	\caption{An additional bottom trapped state in non-percolated CQW on $(2n,0)$ tubes corresponding to the eigenvalue $\lambda = (1-i\sqrt{8})/3$, where $x=-2+i\sqrt{8}$ and $y=1+i\sqrt{8}$ illustrated on an unwrapped tube. The state is localised on the bottom ring of vertices.}
	\label{fig:trapped_iloops_sloops}
\end{figure}

Let us now turn attention to (non-percolated) CQWs. We again stress that any trapped state of PCQW is a trapped state of the corresponding CQW. However,  CQWs can have, in general, some additional trapped states. We briefly discuss their structure. First, for every A-type state in Fig. \ref{fig:trapped_states_both} (a) there is one more  A'-type trapped state for the non-percolated walk corresponding to eigenvalue +1, whose vector elements corresponding to the same undirected edge have both values +1 and -1 as shown in Fig. \ref{fig:trapped_states_both} (b). Second, for chiralities $(2n,0)$ there are also 4 additional trapped states for each eigenvalue $\lambda = (1-i\sqrt{8})/3$ and $\overline{\lambda}$. Due to their degeneracy, we can choose their orthogonal basis in such a way that only the trapped state depicted in Fig. \ref{fig:trapped_iloops_sloops} and its conjugated state have nonzero overlap with the loop initial subspace. We denote both of them as bottom trapped states. Third, there are also additional trapped states which appear only for particular combinations of length and chirality of the tube. Their properties are not well understood yet. However, based on an extensive numerical analysis we conjecture that these trapped states have always non-zero overlap with the loop sink.  It appears that in vertex-to-loops and loops-to-loops regimes these states are not sr-trapped and do not modify the ATP.

The presence of these additional trapped states further modifies transport properties we have found for PCQWs. First of all, ATP of the CQW is never higher then the ATP of the corresponding PCQW, which is known as environment assisted quantum transport \cite{Rebentrost2009,assisted_transport}. Moreover, in  \citep{Kendon2016} authors investigated numerically how long it takes in CQW on a similar nanotube structure, till the walker, initiated in the equal superposition of base states from the vertex subspace, is fully transported to a sink. Our knowledge of trapped states for PCQWs shows that this is the only initial state orthogonal to all these trapped states. Thus, all other choices of the initial state result in nonzero trapping probability in PCQWs and thus also in CQWs. 

\begin{figure}
	\centering
	\includegraphics[width=240 pt]{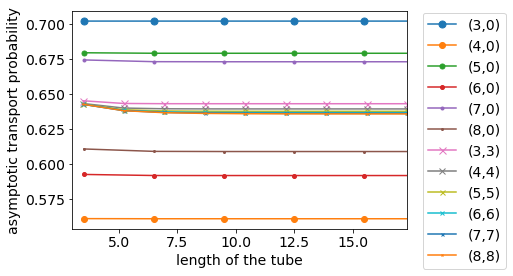}
	\caption{The average ATP for the CQW in the loops-to-loops transport regime for different chiralities and lengths of the tube.}
	\label{fig:unpercolated_loops_loops}
\end{figure}

In our case, all these additional trapped states are orthogonal to trapped states of PCQWs. Thus their contributions to the trapping probability is simply additive. Hence, the loops-to-loops transport regime is the least affected. The only contributing trapped states to the averaged ATP are the two bottom states for chiralities $(2n,0)$. Comparison of Fig. \ref{fig:unpercolated_loops_loops} with Fig. \ref{fig:transport_decrease} demonstrates clearly that the averaged ATP of the CQW and PCQW is the same except for $(2n,0)$ chiralities. This difference between the averaged ATPs can be here evaluated exactly, analogously as in the previous case, using the explicit forms of bottom states. Their overlap with the maximally mixed state on the loops initial subspace is the same and altogether they form the difference in the averaged ATPs as $1/(2\cdot (2n))$, which is in perfect agreement with data from both figures \ref{fig:unpercolated_loops_loops} and  \ref{fig:transport_decrease}. 

\begin{figure}
	\centering
	\includegraphics[width=260 pt]{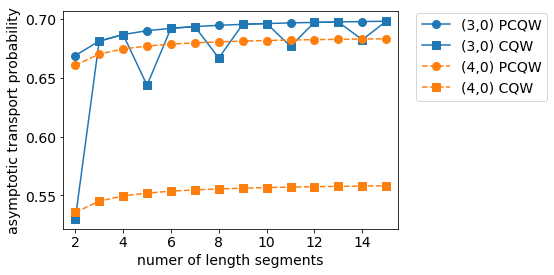}
	\caption{(color online) The average ATP for both PCQW (circles) and CQW (squares) with one-vertex type sink and loops type initial state for $(3,0)$ (blue, solid line) and $(4,0)$ (orange, dashed line) tubes. Since we only present $(n,0)$ tubes, the length is measured in the number of basic length segments for clarity.}
	\label{fig:unperc_sone_iloops}
\end{figure}

A very similar pattern can be observed in the averaged ATP for the CQW in the vertex-to-loops regime. However, as the vertex initial subspace has also overlap with A'-type trapped states, all the averaged ATPs are slightly shifted. On the other hand, we obtain a quite different picture for the CQW in the loops-to-vertex transport regime, see Fig. \ref{fig:unperc_sone_iloops}. Since the loops initial subspace is orthogonal to the A'-type states, the ATP for CQW and PCQW coincides for some lengths of the tube with chiralities $(2n+1,0)$. Sudden drops of the averaged ATP are caused by additional trapped states which appear only for some lengths and chiralities. As expected, for longer tubes the influence of these states on ATP decreases. In contrast, the averaged ATP for CQW on the tube with chirality $(4,0)$ does not exhibit oscillations. Due to the overlap of the loops initial subspace with bottom states the ATP is reduced, compared to the PCQW, regardless of the length of the tube.

As can be seen from figures \ref{fig:unpercolated_loops_loops} and \ref{fig:unperc_sone_iloops}, for non-percolated CQWs the systematic effects of increasing $n$ and length are usually severely disrupted by the presence of additional trapped states. These mostly have significant impact on the ATP, if they are present for the particular choice of the structure parameters. However we can still observe some common features of ATP behavior related to the chirality. First, the highest averaged ATP is in all transport regimes obtained for the thinnest tube $(3,0)$. Second, due to the absence of any additional trapped state, CQWs and PCQWs on tubes with armchair chirality $(n,n)$ have the same ATP in loops-to-loops and vertex-to-loops transport regimes.

In conclusion, we have explored transport properties of Grover coined quantum walks (CQWs) and Grover percolated coined quantum walks (PCQWs) on carbon nanotubes with the armchair $(n,n)$ and the zig-zag $(n,0)$ chirality. Using a general theory for trapped states of PCQWs on planar graphs, we have constructed a convenient basis allowing to uncover how individual geometric characteristics of nanotubes and different types of a sink affect the set of trapped states. Based on this we analyse the asymptotic transport probability (ATP) in dependence of the tube length and chirality for different types of source-to-sink transport regimes. With the analytical insight into the relation between geometric properties of nanotubes and their trapped states, we have found several interesting transport effects. In particular, it is shown for all the studied chiralities that the longer tubes can be surprisingly more efficient for the excitation transport then the shorter ones. In addition, the chirality of the tube is responsible for a quantitative splitting of the average ATP into two main branches, where tubes with the chirality $(n,0)$ show better transport then tubes with the chirality $(n,n)$. We have shown that without a significant effort it is possible to provide a reliable estimate for the averaged ATP on tubes with different chiralities, based solely on a few trapped states having the highest overlap with the initial state. The diameter of tubes further generates a gentle separation for quantitative behavior of the averaged ATP. Quite generally, the thinner tubes exhibit better transport.

A numerical analysis performed for (non-peroclated) CQWs has revealed how a partly known structure of additional trapped states of CQWs further modifies a behavior of the ATP in comparison with PCQWs. It remains true that one can achieve better transport by increasing the length of the tube but for some chiralities this behavior is accompanied with gradually diminishing oscillations. Due to additional trapped states, which appear in CQWs only for some chiralities, the behavior of the averaged ATP splits into more quantitative branches, however, the thinnest tube exhibits the best transport. For a special transport regime we analytically calculated the difference between the averaged ATPs of PCQWs and CQWs.

Finally, let us point out, that we have investigated all phenomena mostly in terms of the averaged ATP. This allows us to make statements about the dominant behavior of the ATP irrespective of the walker's initial state. On the other hand, it also means that for special choices of the initial state these effects are significantly stronger.

{\it Acknowledgements} JM, JN and IJ acknowledge the financial support
from the Czech Science foundation (GA\v CR) project number 16-09824S, M\v{S}MT No. 8J18DE006, RVO14000, Grant Agency of the Czech Technical University in Prague grant No. SGS19/186/OHK4/3T/14, ``Centre for Advanced Applied Sciences'', Registry No. CZ.02.1.01/0.0/0.0/16\_019/0000778, supported by the Operational Programme Research, Development and Education, co-financed by the European Structural and Investment Funds and the state budget of the Czech Republic. IJ is partially supported from GA\v{C}R 17-00844S.

\clearpage


\begin{thebibliography}{99}
\bibitem{review} S. E. Venegas-Andraca, Quantum Information Processing vol. 11(5), pp. 1015-1106 (2012)

\bibitem{Stefanak2008}
M. \v Stefa\v n\'ak, I. Jex, and T. Kiss, Phys. Rev. Lett. {\bf 100}(2), 020501 (2008).

\bibitem{Werner2013}
F. A. Grunbaum, L. Vel\'azquez, A. H. Werner, and R. F. Werner, Commun. Math. Phys. {\bf 320}, 543 (2013).

\bibitem{Tanner2009}
B. Hein and G. Tanner, Phys. Rev. Lett. {\bf 103}, 260501 (2009).

\bibitem{Skoupy2016}
M. \v Stefa\v n\'ak and S. Skoup\'y, Phys. Rev. A {\bf 94}, 022301 (2016).

\bibitem{Aharonov2001}
D. Aharonov, A. Ambainis, J. Kempe, and U. Vazirani, in {\em Proc. of the 33th ACM Symposium on The Theory of Computation, 2001}, (ACM New York, NY, USA, 2001) p. 50.


\bibitem{Magniez}
F. Magniez, A. Nayak, P. Richter, and M. Santha, .Algorithmica {\bf 63}(1), 91 (2012).

\bibitem{Asboth2012}
J. K. Asb\'oth, Phys. Rev. B {\bf 86}, 195414 (2012).



\bibitem{Moore2002}
Ch. Moore and A. Russell, in {\em Proc. of the 6th Intl. Workshop on Randomization and Approximation Techniques in
Computer Science, 2002}, edited by J. D.P. Rolim and S. Vadhan, ( Cambridge, MA, USA, 2002) p. 164.

\bibitem{Portugal2008}
F. L. Marquezino, R. Portugal, G. Abal, and R. Donangelo, Phys. Rev. A \textbf{77}, 042312 (2008).

\bibitem{Segawa2009}
K. Chisaki, M. Hamada, N. Konno, and E. Segawa, Interdisciplinary Information Sciences {\bf 15}, 423 (2009).

\bibitem{Lyu2015}
Ch. Lyu, L. Yu, and S. Wu, Phys. Rev. A {\bf 92}, 052305 (2015).

\bibitem{Kendon2016}
H. Bougroura, H. Aissaoui, N. Chancellor, and V. Kendon, Phys. Rev A {\bf 94}, 062331 (2016).

\bibitem{Segava2013}
N. Konno, N. Obata, and E. Segawa, Commun. Math. Phys. {\bf 322}, 667 (2013).

\bibitem{Lara2013}
P. C. S. Lara, R. Portugal, and S. Boettcher, International Journal of Quantum Information {\bf 11}, 1350069 (2013).


\bibitem{Inui2004}
N. Inui, Y. Konishi, and N. Konno, Phys. Rev. A {\bf 69}, 052323 (2004).

\bibitem{inui:psa}
N. Inui and N. Konno, Physica A \textbf{353} 133 (2005).

\bibitem{inui:grover1}
N. Inui, N. Konno and E. Segawa, Phys. Rev. E \textbf{72} 056112 (2005).

\bibitem{miyazaki}
T. Miyazaki, M. Katori, and N. Konno, Phys. Rev. A {\bf 76} 012332 (2007).

\bibitem{watabe}
K. Watabe, N. Kobayashi, M. Katori and N. Konno, Phys. Rev. A {\bf 77} 062331 (2008).

\bibitem{falkner}
S. Falkner and S. Boettcher, Phys. Rev. A {\bf 90} 012307 (2014).

\bibitem{machida}
T. Machida, Quantum Inf. Comput. {\bf 15} 406 (2015).


\bibitem{Fleischhauer2000}
M. Fleischhauer and M. D. Lukin, Phys. Rev. Lett. {\bf 84}, 5094 (2000).


\bibitem{Poltl2009}
Ch. P\"{o}ltl, C. Emary, and T. Brandes, Phys. Rev. B {\bf 80}, 115313, (2009).


\bibitem{Creatore2013}
C. Creatore, M. A. Parker, S. Emmott, and A. W. Chin, Phys. Rev. Lett. {\bf 111}, 253601 (2013).

\bibitem{Mendoza2013}
J. J. Mendoza-Arenas, T. Grujic, D. Jaksch, and S. R. Clark, S. R. (2013), Phys. Rev. B {\bf 87}(23), 235130 (2013).


\bibitem{Rebentrost2009} P. Rebentrost, M. Mohseni, I. Kassal, S. Lloyd and A. Aspuru-Guzik, New J. Phys. \textbf{11}, 033003 (2009).

\bibitem{Chin2010} A.W. Chin, A. Datta, F. Caruso, S.F. Huelga, and M.B.
Plenio, New J. Phys. {\bf 12}, 065002 (2010).

\bibitem{Caruso2009}
F. Caruso, A. W. Chin, A. Datta, S. F. Huelga, and M. B. Plenio, J. Chem. Phys. {\bf 131}, 105106 (2009).


\bibitem{asymptotic1} B. Koll\'ar, J. Novotn\'y, and I. Jex, Phys. Rev. Lett. {\bf 108}, 230505 (2012).

\bibitem{asymptotic2} B. Koll\'ar, J. Novotn\'y, T. Kiss, and I. Jex, New J. Phys. {\bf 16}, 023002 (2014).

\bibitem{assisted_transport} M. \v Stefa\v n\'ak, J. Novotn\'y, and I. Jex, New J. Phys. \textbf{18}, 023040 (2016).

\bibitem{theory} J. Mare\v s, J. Novotn\'y, and I. Jex, Phys. Rev. A {\bf 99}, 042129 (2019).

\bibitem{transport_effects}
J. Mare\v s, J. Novotn\'y, M. \v Stefa\v n\'ak, and I. Jex, submitted to Phys. Rev. A (2019).


\bibitem{tubegen} TubeGen 3.4 (web-interface, http://turin.nss.udel.edu/research/tubegenonline.html), J. T. Frey and D. J. Doren, University of Delaware, Newark DE, 2011.


\bibitem{Kollar2015}
B. Kollár, T. Kiss, and I. Jex, Phys. Rev. A {\bf 91}, 022308 (2015).















%
%
%
%
%
%
%

%
%


\end{thebibliography}
\end{document}